\pdfoutput=1

\def\year{2019}\relax
\documentclass[letterpaper]{article} 
\usepackage{aaai19}  
\usepackage{times}  
\usepackage{helvet}  
\usepackage{courier}  
\usepackage{url}  
\usepackage{graphicx}  
\usepackage{multirow}
\begin{document}
%
\title{Leveraging Textual Specifications for Grammar-based Fuzzing of Network Protocols}

\author{Samuel Jero\textsuperscript{1}, Maria Leonor Pacheco\textsuperscript{1},
Dan Goldwasser\textsuperscript{1}, Cristina Nita-Rotaru\textsuperscript{2} \\
\textsuperscript{1}Purdue University \\
\textsuperscript{2}Northeastern University\\
\{sjero, pachecog, dgoldwas\}@purdue.edu\\
c.nitarotaru@neu.edu}

\maketitle
\begin{abstract}

Grammar-based fuzzing is a technique used to find software vulnerabilities by injecting well-formed inputs generated following rules that encode 
application semantics. 
Most grammar-based fuzzers for network protocols rely on human experts to  manually specify these rules.
In this work we study automated learning of protocol rules from textual specifications (i.e. RFCs).
We evaluate the automatically extracted protocol rules by applying them to a state-of-the-art fuzzer for transport protocols and show that it leads to a smaller number of test cases while finding the same attacks as the system that uses manually specified rules.
\end{abstract}

Ensuring that protocol implementations are free of bugs and vulnerabilities 
is an important problem, given the reliance of virtually any application on computer networks. One testing approach used to tackle this problem is protocol fuzzing, which generates and injects packets into the protocol stream. To successfully apply this method and increase the likelihood of finding protocol vulnerabilities, the generated packets need to be crafted carefully. A technique that has been shown to be effective in creating these attack packets is grammar-based fuzzing~\cite{snake_dsn_2015,Abdelnur2007,Banks2006,Wang2013,jero2017beads}. In this technique, attack packets are generated following rules that encode protocol semantics. For example, in the TCP protocol, bytes 17 and 18 contain a checksum of the rest of the TCP header. An injected packet must contain the correct checksum in order to pass the trivial checksum check and reach the part of the code that it intends to test. 

The effectiveness of grammar-based protocol fuzzers depends significantly on the protocol rules they use. Unfortunately, these rules are usually created manually by an expert and are not easily transferable from one protocol to another. Our main observation is that there is an untapped resource of information available for network protocols in the form of natural language specification documents (e.g. RFCs).
With the recent interest in using data to solve problems in several fields, we ask the question: ``\emph{Can we leverage natural language specifications of protocols to improve protocol fuzzers?}" 

 In this paper, we study how to improve the coverage and effectiveness of grammar-based fuzzers for network protocols through automated learning of protocol rules from existing textual documentation. We have two design goals: (1) minimize the manual supervision effort required for training and (2) adapt to new protocols without re-training.
  
\begin{figure}[b]
  \vspace{-20pt}
	\centering
	\includegraphics[width=.74\linewidth]{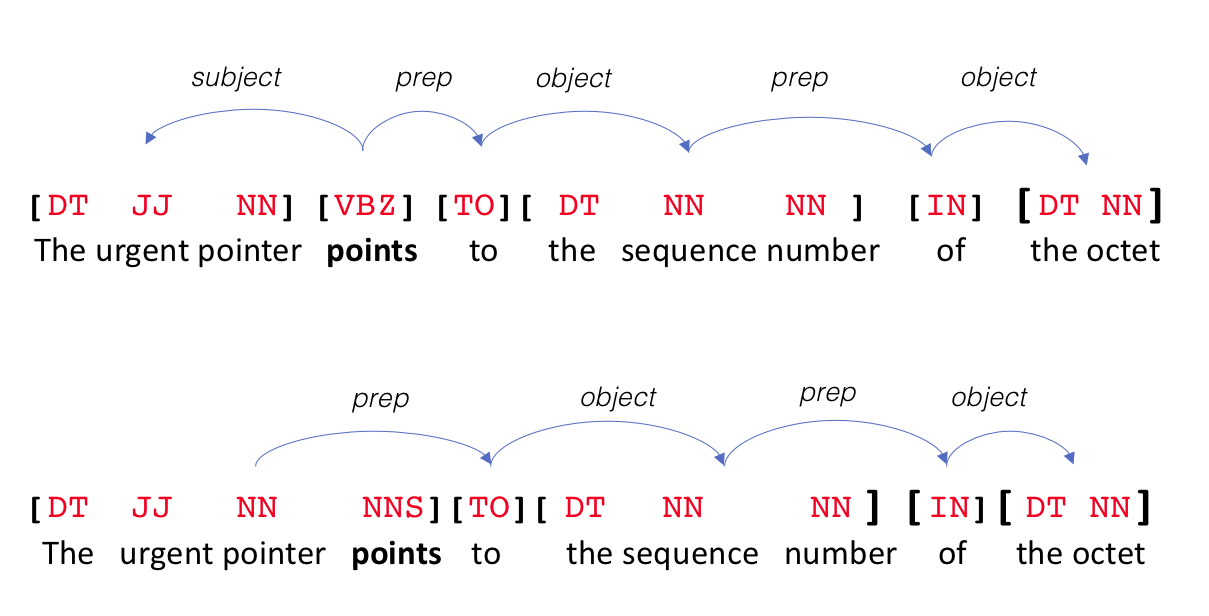}
		\vspace{-10pt}
	\caption{NLP analysis: two possible outcomes for the senses of the word \textit{``points''} (top: verb, bottom: noun). }
	\vspace{-10pt}
	\label{fig:nlp}
\end{figure}

  
Extracting protocol information is often not straight-forward. Protocol specifications intended for human readers capable of understanding context and intent are often too ambiguous for simple rules-based extraction. Even relying on the recent advances in NLP technology by applying ``off-the-shelf" NLP tools can result in reduced performance and brittle applications due to domain differences. The performance of these tools, typically trained on newswire data, drops significantly when applied to technical specification documents.   
Figure~\ref{fig:nlp} demonstrates this 
 point, showing a misclassification of the word ``\textit{points}'' as a noun, common in the newswire data used for training, but incorrect in the given context. Note that this mistake propagates to other steps in the pipeline, resulting in incorrect chunking and parsing decisions. Our main technical challenge is to adapt information extraction to the new domain without incurring the high cost of providing supervision for each individual protocol.

Our contributions are as follows: (1) \textit{We define the problem of protocol grammar extraction as a set of NLP tasks}. The grammar is designed to capture relevant aspects and consists of the protocol's header fields and their properties.
%
(2) \textit{We suggest and evaluate an NLP framework for these tasks, designed to reduce manual supervision effort when adapting to new network protocols.}
Unlike previous work that applied transformation rules to the output of NLP tools directly, we propose a lightweight zero-shot learning framework which can adapt to the specific properties of the networking domain. 
%
 (3) \textit{We demonstrate the usefulness of the information extracted by our NLP framework by applying it to a transport protocol fuzzer}.
 We compare its performance, when using manually and automatically extracted protocol information, on two transport protocols and find that our automatically generated protocol grammars are as effective in identifying attacks as manually created grammars while often enabling improved efficiency.




\section{Related Work}
\label{sec:related}

Previous work has applied  NLP techniques to related problems. WHYPER~\cite{whyper} and DASE~\cite{dase} apply NLP techniques to identify sentences that describe the need for a given permission in a mobile application description and extract command-line input constraints from manual pages, respectively. The work in~\cite{TM} used documentation and source code to create an ontology allowing the cross-linking of software artifacts represented in code and natural language on a semantic level. These approaches focus on a small, predefined set of entities; analyze small, structured sentences; and use rule-based approaches.
Other works infer protocol specifications using network traces \cite{comparetti2009prospex,wang2011inferring,cho2010inference}, program analysis~\cite{kothari2008deriving,Cho2011,lin2008automatic}, or model checking~\cite{lie2001simple,corbett2000bandera}. These approaches rely extensively on input from human experts and do not easily generalize to new software or protocols.



\section{Problem Definition}
\label{sec:background}



\paragraph{Protocol Grammar-based Fuzzing}
\begin{figure}
	\centering
	\includegraphics[width=0.75\linewidth]{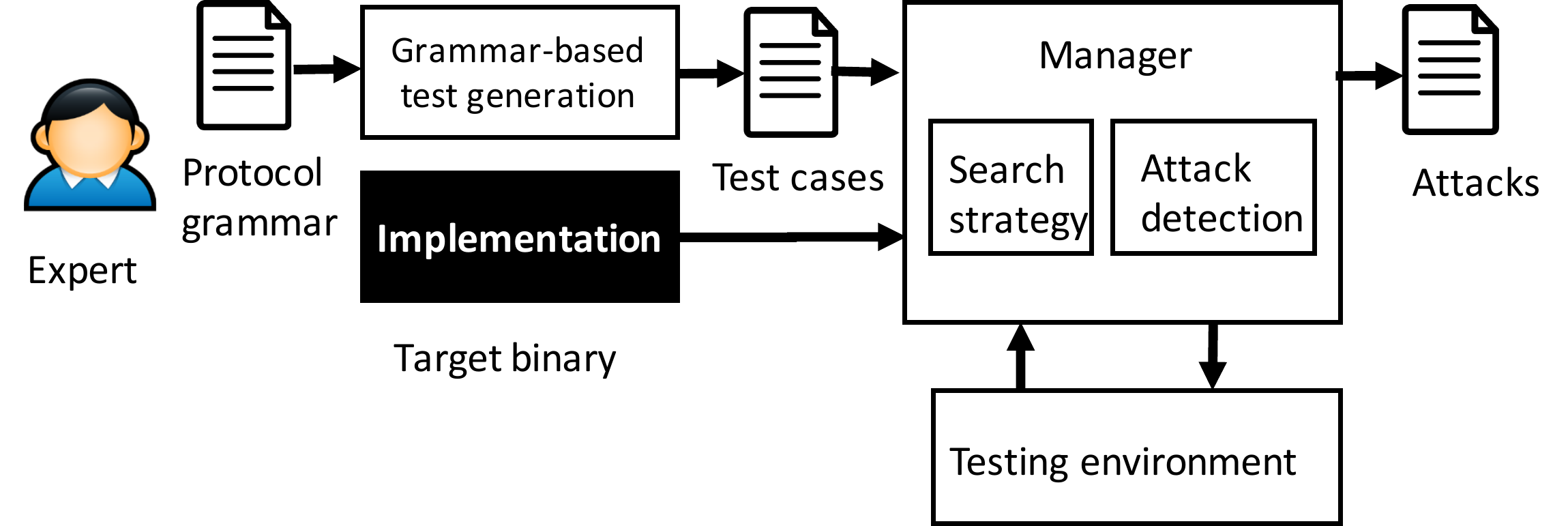}
	\vspace{-5pt}
	\caption{Grammar-based fuzzing.}
	\label{fig:fuzzing}
		\vspace{-5pt}
\end{figure}
Fuzzing is a technique used for finding software vulnerabilities by injecting random inputs and then observing the output of the program under test. In grammar-based fuzzing the injected inputs follow rules encoding relevant system properties. For network protocols, inputs consist of packets, and rules represent protocol semantics, such as  
properties-of, and relations-between, packet header fields.
%
Fig.~\ref{fig:fuzzing} shows an overview of the fuzzing process where an expert
manually specifies the rules used to generate testing strategies.
A manager script takes the test cases and tested system as inputs
and interacts with a testing environment 
to instantiate the tested protocol, inject the test cases,
and receive reports about the execution. 

%
	\vspace{-12pt}
\paragraph{Protocol Grammar Extraction}
A network protocol is defined by the header attached to transported packets. This header
often has fixed size (in bits), where certain parts of it, known as fields, have defined meaning and size. Protocol semantics are defined by the 
properties of these fields. E.g., bytes 17-18 in the TCP header contain a checksum.

We formalize these concepts and define protocol grammars over two components:
a set of \textit{fields} that 
correspond to the header, with each field having a name, a size (i.e. \# bits), and an 
order in the packet header, as well as a set of optional field \textit{properties}.
These definitions are given with respect to a set of protocol-specific named fields ($f$) and field-specific named properties ({\small$\langle f,p\rangle$}). 
%
%
%
Given these notations, we define two NLP tasks for extracting protocol information. 
 The first, \textbf{Type Extraction}, given a protocol document, extract the set of protocol field and property symbols.
  The second problem, \textbf{Symbol Identification and Linking}, given the document and the set of extracted symbols, identify mentions of these symbols in text, and link together field mentions to their relevant properties,
 as indicated by the protocol text.

 \vspace{-10pt}
\paragraph{Zero Shot Learning for Entity and Property Linking.} 
Approaching these problems using a traditional fully supervised approach would require building a separate classifier for each specific set of protocol symbols, \texttt{T $\rightarrow$E}, which would defeat the goal of automating the process. 
Instead, we take a  zero-shot learning (ZSL)~\cite{palatucci2009zero} approach, which learns a mapping \texttt{$\langle$T,E$\rangle\rightarrow$ \{t,f\}} from a tuple containing the input and output to a Boolean value indicating whether the pair is correct or not. The main observation behind zero-shot learning is that the set of output symbols does not have to be fully specified during training, and unlike traditional supervised learning, \textit{the system is expected to perform well even over outputs that were not observed during training}. This is done by learning a similarity metric, $sim(t,e_i)$ and defining the prediction as: $\arg\max_{e_i\in E} sim(t,e_i)$. We learn a similarity function between textual phrases and protocol fields and 
properties. The similarity function captures the surface level string similarity, acronyms used in the text to refer to the fields, and anaphoric references (``it'',``that field'') based on their context. This approach adapts to new protocols by providing different sets of entities. 


\section{Design}\label{sec:design}

 We designed the NLP pipeline to solve the two problems above with two goals: (1) minimize the manual supervision effort required for training and (2) adapt to new protocols without re-training the system. 
First, we include a pre-processing step to read in the raw specification documents and normalize their structure. Then, the entity types extraction task leverages the hierarchical structure of protocol specification documents, like RFCs.
We use a rule-based system leveraging RFC specific formatting for identifying and extracting entity types. There are 25 types for each protocol on average, and the rule based system was able to recover these types with 0.82 accuracy. We limit the discussion due to space consideration.

We take a two-step approach for the symbol identification task, first locating field (entity) mentions in the document, and then, by examining their context, we look for properties 
 associated with them. For both steps we use a ZSL approach, where
a classifier is trained to look for similarities between document text and protocol symbols. 
%
We developed new classifiers trained on network protocol data instead of using off-the-shelf tools trained for non-technical domains, which are a  poor fit for our highly technical domain.

Finally, a post-processing step transforms the information extracted into a protocol grammar description, which can be used by downstream packet generation tasks.

\begin{figure}
	\centering
	\includegraphics[width=0.7\linewidth]{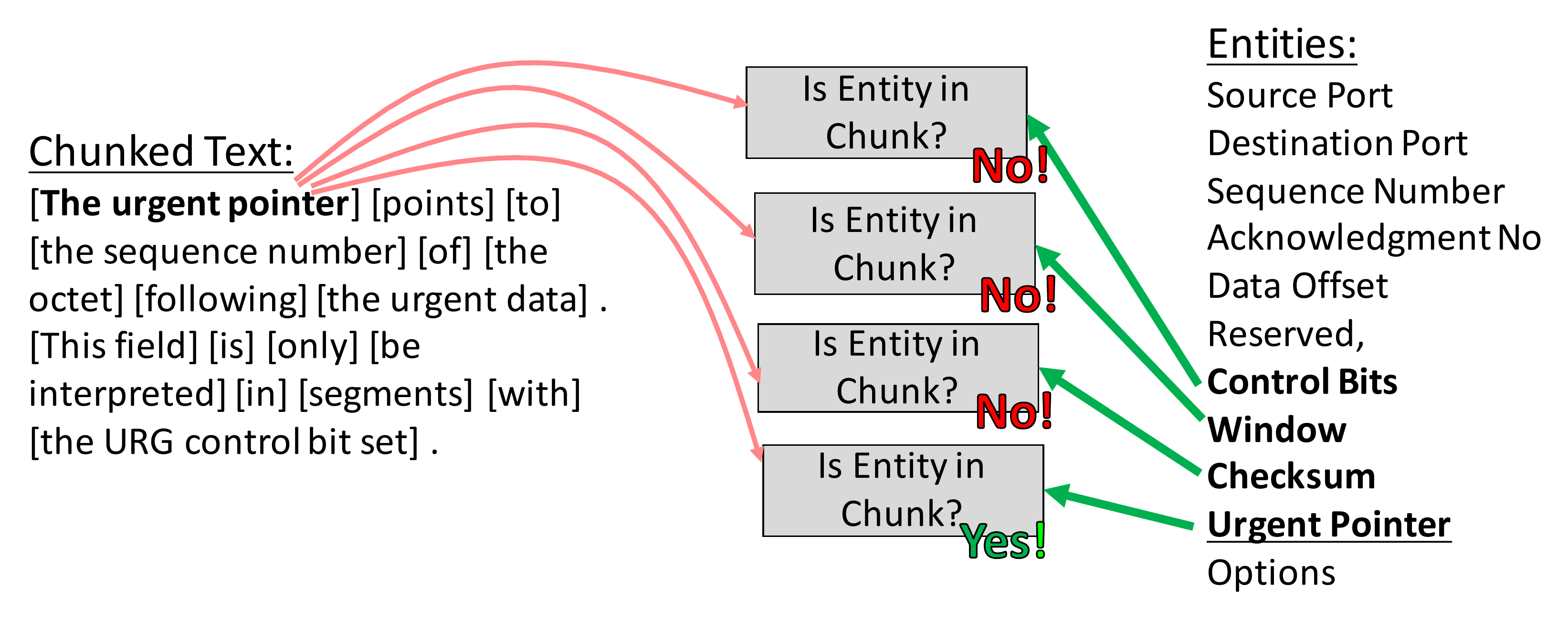}
	\caption{Example of zero-shot learning classification for entity mentions.}
	\vspace{-10pt}
	\label{fig:classifiers}
\end{figure}
 
 \vspace{-12pt}
\paragraph{Entity Mention Identification}
For this task, the needed inputs are the pre-processed document and the list of entity types. We used the entity types list we extracted automatically from each document,
but any ontology consisting of relevant entity types could also be used. 
Since entity types vary in both name and number between protocols, we use a ZSL approach 
that learns a similarity metric between text snippets that takes into account character level similarity, writing style (e.g., capitalization patterns, abbreviations), and relevant context words.
This approach allows our classifier to {\em  generalize to previously unseen entity types} that appear in new protocol documents.

Specifically, we define a binary classification problem over all pairs $(e_j, c_i)$ where $e_j$ represents each entity type and $c_i$ represents a chunk in the document text, as shown in Fig.~\ref{fig:classifiers}. If the chunk $c_i$ contains a reference to entity type $e_j$, the pair is labeled as a positive example. The pair is labeled as a negative example otherwise. This way we learn a similarity score between $e_j$ and $c_i$ that is able to generalize to different entity types.
We train an SVM classifier for this problem using a set of binary features (description omitted due to space, we will release our feature extraction code). 


\begin{figure*}{\small
	\raggedright
	{\bf Property extraction:}\\
	Section Title: [(entity mention: Data Offset)Data Offset] : 8 bits\\
	Section Text: The offset from the start of the packet 's DCCP [(property keyword: header length) {\em header}] to the start of its application data area , in 32-bit words . The receiver must ignore packets whose Data Offset is smaller than the minimum-sized header for the given Type or larger than the DCCP packet itself .\\
	{\em Property: Header Length, Data Offset}}\\
		\vspace{-20pt}
	\caption{Example of property extraction.}
	\label{fig:example_pr}
	\vspace{-10pt}
\end{figure*}

\vspace{-12pt}
\paragraph{Property Extraction}
At this stage we identify the properties of 
entity types and extract them from the document body.
We used the ZSL approach in this case as well. Based on an analysis of a wide variety of network protocols, we selected 9 properties to extract.
The properties we consider include \texttt{checksum}, which marks packet fields containing checksums; \texttt{port}, which marks packet fields used for multiplexing different communication channels; and \texttt{multiple}, which indicates that a field's value is a multiple of some constant. 

Note that unlike entity types, which vary between protocols, we look for the same properties 
in each protocol. We choose these properties 
because they are widely present across network protocols and contain information that is useful for generating test cases. For example, knowing that a field represents a checksum means that we should not spend a lot of time testing random values for that field. 


This classifier identifies chunks of text that express 
a property; however, it does not determine which property
nor the identity of the arguments (i.e. the entity types involved in the relation or property). 
Identifying the type of a relation or property is done simply by choosing the relation or property with the maximum key phrase overlap. To determine the argument of the property,
we use the entity mentions identified in the previous stage and a simple heuristic.
We choose the entity type defined in the title of the section in which the property appears. Since many properties refer to the entity type currently being discussed, this makes sense. 
Fig.~\ref{fig:example_pr} shows an example of the output of this classifier.

\vspace{-12pt}
\paragraph{Post-processing}
%
%
We post-process the properties 
by leveraging domain specific knowledge.
Since these properties 
are being used to characterize the protocol, we only need a single $(property,entity)$ 
tuple, regardless of how many times it appears in the document. This benefits us significantly as we usually have multiple opportunities to extract each property tuple.
In addition, many properties can occur only on a single field in the packet header (e.g. \texttt{packet type}, \texttt{header length}), while others 
cannot occur in combination (e.g. \texttt{packet type} and \texttt{sequence number} are mutually exclusive). Finally, if our pipeline was unable to identify key properties like \texttt{packet type}, \texttt{header length}, and \texttt{checksum}, we attempt to guess which fields have these properties based on field names and sizes. Finally, we associate our cleaned properties 
with the packet fields.


\section{Evaluation}
\label{sec:eval}

We do an intrinsic evaluation of our system, observing the performance of the ZSL setup at extracting entities and properties for different protocols. Then, we do an extrinsic evaluation, using the extracted information as an input to a grammar-based fuzzer for transport protocols.

\subsection{Information Extraction Evaluation}


\paragraph{Zero-Shot Learning Setup}
We formalize the ZSL setup as follows. Given a set of $N$ RFC documents, each describing a different network protocol, we learn scoring functions for extracting entities and properties using $N-1$ documents, and test on a different document with unobserved entity types. Each document $D_i$ has a specific set of entity types $E_i$. As mentioned in the design section, the set of properties {\small $P = \{p_{1}, p_{2}, ..., p_{m} \}$} is the same for every protocol. 

We define a training set $S = (D,E)$, where $D$ is a set of $N-1$ documents and $E$ is the set of entity types extracted from those documents. Then, we learn a scoring function $s_1(e_j, c_i)$ to determine the likelihood of a chunk $c_i \in D_i$ of being a mention of an entity $e_{j} \in E_i$. In the case of properties, the amount of training data is very limited. For this reason, we generalize the problem and learn a scoring function $s_2(c_i, P)$ to determine if a chunk $c_i \in D_i$ is a mention to \textit{any} property in $P$. We solve the problem of identifying the property type by selecting the type with maximum key phrase overlap with the chunk. 

At test time, the aim is to take a new document $D_k$, with unobserved entity types $E_k$, and extract properties and entities from its specification. Once mentions are extracted, we match each property with the entity defined in the title of the section in which the property appears. 

\vspace{-12pt}
\paragraph{Dataset}
We annotated a set of public RFC documents.\footnote{https://www.ietf.org/standards/rfcs/} These documents are a common form for protocol specification and are written in plain text following a specific format. We use RFC documents for six protocols: GRE, IPv6, IP, TCP, DCCP and SCTP. 

\vspace{-12pt}
\paragraph{Results}
For entity mentions, we measure precision, recall, and F1. Additionally, we report the number of true positives (TP) and false positives (FP). In the case of properties, annotations often span more than one chunk. We consider a property to be found if we classify any chunk in the annotation span as a property. For this reason, we report the true positive rate at the span level (S-TPR). Additionally, we measure the rate of false chunks that our classifier miss-classifies as properties (C-FPR). 

	\begin{table}{}
		\begin{center}
		\small
			\caption{Entity Mention Identification}
			\label{table:reference_baselines}
			\begin{tabular}{|l||l|l|l|l|l|}
				\hline
				Approach & Prec & Recall & F1 & TP & FP \\
				\hline		
				O $\geq$ 50\%  & 0.19 & 0.58 & 0.29 & 502 & 2147\\
				O $\geq$ 70\%  & 0.40 & 0.48 & 0.43 & 418 & 617\\
				O $\geq$ 85\%  & 0.58 & 0.42 & 0.49 & 363 & 258\\
				O $\geq$ 100\% & 0.74 & 0.36 & 0.49 & 316 & 111\\
				$RB_1$               & 0.93 & 0.18 & 0.30 & 157  & 12\\
				$RB_2$               & 0.77 & 0.48 & 0.59 & 411 & 122\\
				\textbf{Our Approach}& \textbf{0.78} & \textbf{0.66} & \textbf{0.72} & \textbf{576} & \textbf{159} \\
				\hline
			\end{tabular}
		\end{center}
				\vspace{-5pt}
	\end{table}
	\begin{table}{}
		\begin{center}\small
			\caption{Property Extraction}
			\label{table:relations_baselines}
			\begin{tabular}{ |l|l|l|} 
				\hline
				Approach &  S-TPR &  C-FPR  \\
				\hline
				O $\geq$ 50\% &  0.86 & 0.36 \\  
				O $\geq$ 70\% &  0.77 & 0.12  \\  
				O $\geq$ 85\% &  0.77 & 0.11 \\  
				O $\geq$ 100\% &  0.77 & 0.11 \\  
				$RB_1$ &  0.90 & 0.89\\  		
				$RB_2$ &  0.95 & 0.87\\  
				\textbf{Our Approach} &  \textbf{0.86} & \textbf{0.28}   \\
				\hline
			\end{tabular}
		\end{center}
				\vspace{-5pt}
	\end{table}

We do six iterations, training with five protocols and testing on the sixth. Tables \ref{table:reference_baselines} and \ref{table:relations_baselines} show aggregated results for these six iterations. We compare our approach with a set of rule based systems. 

Table \ref{table:reference_baselines} shows results for extracting entity mentions. The first four rows correspond to simple string matching systems. Here, we measure the overlap $O$ between an entity type and the current chunk. We classify the chunk as an entity mention if the overlap is at or above a certain percentage $P$. The trade-off in these systems is clear. The higher $P$, the higher the precision and the lower the recall. As we reduce $P$, recall increases and precision suffers. The following two approaches are rule-based systems based on our feature set. Here, we take the same set of features used by our classifier and weigh them manually. In $RB_1$, we weight each feature by its frequency of occurrence in the dataset. For each feature $f$ we calculate $pr$ and $nr$. We then give each feature $f$ a weight of $+pr$ if $pr > nr$, a weight of $-nr$ if $nr > pr$, and a weight of 0 if $pr = nr$. We use a weight of $-nr$ for the bias term. In $RB_2$, we weight each feature with $+1$ if it occurs more often in positive examples and $-1$ if it occurs more often in negative examples. We use a weight of $-1$ for the bias term. While $RB_2$ performs better than string matching, our classifier still outperforms all baselines. In other words, there is value in both informative features and the use of our learning framework. 

	\begin{table}{}
		\begin{center}\small
			\caption{Entity Mention Identification per Protocol}
			\label{table:references}
			\begin{tabular}{|l||l|l|l||l|}
				\hline
				Protocol & Prec & Recall & F1 & \# Inst  \\
				\hline	
				TCP  & 0.96 & 0.68 & 0.80 & 38\\
				SCTP & 0.70 & 0.60 & 0.64 & 484 \\
				IPv & 0.93 & 0.80 & 0.86 & 127 \\
				IP & 0.87 & 0.60 & 0.71 & 45 \\
				GRE & 1.0  & 0.81 & 0.89 & 21 \\
				DCCP & 0.85 & 0.73 & 0.79 & 160 \\
				\hline
				\textbf{Total (K)} & \textbf{0.78}  &  \textbf{0.66} & \textbf{0.72}   & \textbf{875} \\ \hline
				\textbf{Total (E)}  &  \textbf{0.73} & \textbf{0.53} & \textbf{0.62} & \textbf{875}  \\  
				\hline
			\end{tabular}
			\vspace{-15pt}
		\end{center}
	\end{table}

For properties, results can be observed in Table \ref{table:relations_baselines}. Similarly to the entity mention case, in the first four approaches, we measure the overlap between property key phrases and the current chunk. We classify a chunk as a property if the overlap is at or above a certain percentage $P$. These methods have a high success rate (S-TPR) while introducing less noise (C-FPR). However, the C-FPR is too high for $O \geq 50$ and the S-TPR is too low for $O \geq 70$. Identifying most properties is essential for the performance of the fuzzer, while we can live with \textit{some} level of noise and rely on the post-processing step. We find that our approach gives us a better balance between the number of properties found and the level of noise introduced. $RB1$ and $RB2$ are the same rule-based methods that we considered for entity mention identification. In this case, the level of noise introduced with these systems is too high. 

On table \ref{table:references} we can see the results for extracting entity mentions by protocol. We show that our ZSL approach generalizes well to different, unobserved protocols. We report aggregated results both assuming that the list of entity types is \textit{known} a priori (K), and when the list of entity types is \textit{extracted} using the RFC format (E). Even though performance suffers, we only need to identify a single $(property,entity)$ tuple, regardless of how many times it appears in the document, to leverage this information in the fuzzer. For this reason, the error propagation when using a fully automated pipeline is minimized. Due to space considerations, we only show results by protocol for entity mentions.

\subsection{Fuzzer Evaluation}\label{section:attack_discovery}

\begin{table*}
	\begin{center}\small
		\caption{Coverage Evaluation}
		\label{table:coverage}
		\begin{tabular}{ |l||p{1.7cm}|p{1.5cm}||p{1.7cm}|p{1.5cm}|} 
			\hline
			~ & \multicolumn{2}{c||}{TCP} &\multicolumn{2}{c|}{DCCP}\\
			~ & Unique Pkt Type Traces & Total Strategies & Unique Pkt Type Traces & Total Strategies\\
			\hline
			Random                       &  13  & 1000 &  18 & 1000\\
			Manual						 & 784  & 901  & 718 &  871\\
			NLP-based					 & 713  & 819  & 816 & 1022\\
			\hline
		\end{tabular}
	\end{center}
			\vspace{-5pt}
\end{table*}

\begin{table*}
	\begin{center}\small
		\caption{Attack Discovery Results}
		\label{table:attackdiscovery}
		\begin{tabular}{ |l||p{1.5cm}|p{1.5cm}|p{1.25cm}||p{1.5cm}|p{1.5cm}|p{1.25cm}|} 
			\hline
			~ & \multicolumn{3}{c||}{TCP} &\multicolumn{3}{c|}{DCCP}\\
			~ & Reported Attacks & Interesting \newline (Off-path) Attacks & Unique Attacks & Reported Attacks & Interesting \newline (Off-path) Attacks & Unique Attacks\\
			\hline
			Random                       & 996  & 0  & 0 & 992 &  0 & 0\\
			Manual						 & 219  & 63 & 5 & 209 & 44 & 2\\
			NLP-based					 & 220  & 69 & 5 & 254 & 47 & 2\\
			\hline
		\end{tabular}
	\end{center}
			\vspace{-5pt}
\end{table*}


\paragraph{SNAKE Fuzzer}
We demonstrate the usefulness and effectiveness of our automated protocol grammar extraction framework by applying it to SNAKE~\cite{snake_dsn_2015}, a state-of-the-art transport protocol fuzzer. 
The key component of SNAKE is a malicious proxy that modifies and injects attack packets based on a protocol description manually specified by an expert. 

\vspace{-12pt}
\paragraph{Fuzzer configurations}
We use SNAKE to test two protocols, TCP and DCCP, in a single operating system, Linux 3.0.0 in Ubuntu 11.10. We compare three different testing configurations: Random, Manual, and NLP-based.

{\bf Random.} This configuration uses a fuzzer configured with no information about the protocol grammar. It generates tests that randomly replace a random number of the first 20 bytes of packets with random data. We only modify the first 20 bytes to approximate the length of a typical transport protocol header. Note that in any given test, the same bytes in all packets are modified. Attack injection is on every packet sent. We generate 1,000 test strategies in this manner to compare with our other testing configurations.

{\bf Manual.} This configuration uses the SNAKE fuzzer with a manually created protocol grammar.
 For each packet type, test strategies are created to inject new messages, modify all packet fields, and apply all delivery actions to those packets. For modifying packet fields, tests modify fields based on their size. Attack injection is on every sent packet.


{\bf NLP-based.} This configuration uses SNAKE configured with our automatically extracted protocol grammar, derived from extracted entities and properties. This configuration generates a similar set of tests that injects new packets, modifies the delivery of packets, or overwrites a single field in packets during each test. During each test, all packets of a particular type are modified, and attack injection is on every packet. For each packet type, test strategies are created to inject new messages, modify all packet fields, and apply all delivery actions to those packets. This configuration has more information about packet fields available to it, thanks to our pipeline. We leverage this information to apply better field modifications. For example, from the definition of checksums and protocol ports, we expect that tampering with them will result in modified packets simply being thrown away. Thus, we can apply a single modification to fields that are identified as checksums or ports.

\vspace{-12pt}
\paragraph{Metrics}
To evaluate the different configurations we focus on a number of indicators: (1) the amount of \emph{effort} required to test an implementation; (2) the \emph{coverage} of the generated tests; and (3) the overall \emph{attack discovery} results. 

We use the number of test strategies generated to measure the amount of effort required to test an implementation. We measure coverage as the number of unique packet type traces observed. A \emph{packet type trace} records the order in which different types of packets are observed in a flow. Thus, a packet type trace succinctly summarizes a protocol connection and approximates the path traversed through the code. To effectively test a protocol, as many \emph{unique} connections, or code paths, as possible should be explored.\footnote{Note that we record packets prior to any possible modification to avoid counting traces where the only different packet is one that was intentionally modified.} Ideally, we want to expend a small amount of effort while achieving high coverage. These indicators are reported in Table \ref{table:coverage}. 

The number of attacks identified indicates how many test strategies were reported by the testing configuration as attacks. Unfortunately, many of these attacks are on-path attacks which are not \textit{interesting} (i.e., relevant) since TCP and DCCP do not attempt to protect against these attacks. Removing these on-path attacks leaves us with the interesting off-path attacks, which we refer to as \emph{interesting attacks}. Note that many strategies may exercise the same underlying root vulnerability, so we perform a manual analysis of all reported attack strategies to identify the number of \emph{unique attacks} actually identified. Attacks are reported in table \ref{table:attackdiscovery}.


\vspace{-12pt}
\paragraph{Random Testing vs Grammar-based Fuzzing}
Table~\ref{table:coverage} compares coverage, in terms of unique packet type traces, achieved by all three configurations. We observe that the manual and NLP-based configurations achieve similar coverage, around 700 unique traces for either protocol, while random achieves only 13 traces for TCP and 18 for DCCP. To achieve this coverage, all three configurations required about 1,000 strategies. Since number of strategies is directly equivalent to the amount of effort required for testing, we can say that random testing is significantly less efficient than grammar-based fuzzing.

This occurs primarily because in the random test configuration all packet manipulation strategies stall the connection, since modifying the packet corrupts the protocol checksum, resulting in the packet being thrown away at the receiver. In order to correct this, the fuzzer would need to know the exact location of the checksum in the packet, which is exactly the information provided by a protocol grammar. Similarly, all packet delivery strategies in the random test configuration stall the connection because they drop or delay key packets like the TCP SYN and the DCCP Request. In order to work around this, the fuzzer would need to know the type of each packet, which is also supplied by a protocol grammar. All of these connection stalls generate similar traces and traverse similar code paths, resulting in very poor coverage.

In addition to poor coverage, Table~\ref{table:attackdiscovery} indicates that the random test configuration also generates a significant amount of reported attacks, but none of them are interesting. This is because each of the connection stalls mentioned above is reported as an attack on availability. Unfortunately, these are on-path attacks, not relevant for TCP or DCCP.

\vspace{-12pt}
\paragraph{NLP-based vs Manual Configurations}
We first consider testing coverage, shown in Table~\ref{table:coverage}, and confirm that, thanks to the additional properties provided by our document processing pipeline, the NLP-based configuration generates fewer strategies than the manual configuration for TCP. This results in a reduction in the amount of time and effort required for testing. This does result in slightly lower coverage, but only by about 70 traces.

Unfortunately, for DCCP our pipeline over-approximates the number of fields in each packet, due to differences between packet types. This leads to generating more strategies (1022 instead of 871) and an overall increase in the time and computational effort required for testing. Note that it also results in improved coverage by almost 100 traces.

In terms of the attacks that are reported by our testing configurations, shown in Table~\ref{table:attackdiscovery}, we find that our NLP-based testing system reports a few more attacks (1 more for TCP and 45 more for DCCP) and that more of those reported attacks are interesting. 


 \section{Conclusion}
\label{sec:conc}
In this work, we proposed a methodology for information extraction for technical documents designed around the issues of domain adaptation and minimal supervision, which are repeating issues when using NLP in technical domains. 
We build an NLP framework to extract grammars from natural language specification documents automatically and combine it with a grammar-based fuzzer to create a completely automated testing system. Our document processing pipeline extracts protocol entity types and mentions---or packet fields--- and properties from natural language network protocol RFCs using a zero-shot learning approach. 
We demonstrate the value of our approach by applying it to a transport protocol fuzzer and comparing it to using a manual grammar on two protocols, TCP and DCCP. We find a reduction in the testing effort for TCP, while identifying the same set of attacks, and doing so in a fully automated manner for both TCP and DCCP.


\bibliographystyle{aaai}
\bibliography{mybib}

\begin{thebibliography}{}

\bibitem[\protect\citeauthoryear{Abdelnur, State, and
  Festor}{2007}]{Abdelnur2007}
Abdelnur, H.~J.; State, R.; and Festor, O.
\newblock 2007.
\newblock {KiF}: A stateful {SIP} fuzzer.
\newblock In {\em ACM IPTComm}.

\bibitem[\protect\citeauthoryear{Banks \bgroup et al\mbox.\egroup
  }{2006}]{Banks2006}
Banks, G.; Cova, M.; Felmetsger, V.; Almeroth, K.; Kemmer, R.; and Vigna, G.
\newblock 2006.
\newblock {SNOOZE}: Toward a {S}tateful {N}etw{O}rk pr{O}tocol fuz{ZE}r.
\newblock In {\em ISC}.

\bibitem[\protect\citeauthoryear{Cho \bgroup et al\mbox.\egroup
  }{2010}]{cho2010inference}
Cho, C.~Y.; Shin, E. C.~R.; Song, D.; et~al.
\newblock 2010.
\newblock Inference and analysis of formal models of botnet command and control
  protocols.
\newblock In {\em ACM CCS}.

\bibitem[\protect\citeauthoryear{Cho \bgroup et al\mbox.\egroup
  }{2011}]{Cho2011}
Cho, C.~Y.; Babic, D.; Poosankam, P.; Chen, K.~Z.; Wu, E.~X.; and Song, D.
\newblock 2011.
\newblock {MACE}: Model-inference-assisted concolic exploration for protocol
  and vulnerability discovery.
\newblock In {\em USENIX Security}.

\bibitem[\protect\citeauthoryear{Comparetti \bgroup et al\mbox.\egroup
  }{2009}]{comparetti2009prospex}
Comparetti, P.~M.; Wondracek, G.; Kruegel, C.; and Kirda, E.
\newblock 2009.
\newblock Prospex: Protocol specification extraction.
\newblock In {\em IEEE SP}.

\bibitem[\protect\citeauthoryear{Corbett \bgroup et al\mbox.\egroup
  }{2000}]{corbett2000bandera}
Corbett, J.~C.; Dwyer, M.~B.; Hatcliff, J.; Laubach, S.; P{\u{a}}s{\u{a}}reanu,
  C.~S.; Bby, R.; and Zheng, H.
\newblock 2000.
\newblock Bandera: Extracting finite-state models from java source code.
\newblock In {\em ACM/IEEE ICSE}.

\bibitem[\protect\citeauthoryear{Jero \bgroup et al\mbox.\egroup
  }{2017}]{jero2017beads}
Jero, S.; Bu, X.; Nita-Rotaru, C.; Okhravi, H.; Skowyra, R.; and Fahmy, S.
\newblock 2017.
\newblock {BEADS:} automated attack discovery in {OpenFlow}-based {SDN}
  systems.
\newblock In {\em RAID}.

\bibitem[\protect\citeauthoryear{Jero, Lee, and
  Nita-Rotaru}{2015}]{snake_dsn_2015}
Jero, S.; Lee, H.; and Nita-Rotaru, C.
\newblock 2015.
\newblock Leveraging state information for automated attack discovery in
  transport protocol implementations.
\newblock In {\em IEEE/IFIP DSN}.

\bibitem[\protect\citeauthoryear{Kothari, Millstein, and
  Govindan}{2008}]{kothari2008deriving}
Kothari, N.; Millstein, T.; and Govindan, R.
\newblock 2008.
\newblock Deriving state machines from tinyos programs using symbolic
  execution.
\newblock In {\em IPSN}.

\bibitem[\protect\citeauthoryear{Lie \bgroup et al\mbox.\egroup
  }{2001}]{lie2001simple}
Lie, D.; Chou, A.; Engler, D.; and Dill, D.~L.
\newblock 2001.
\newblock A simple method for extracting models from protocol code.
\newblock In {\em IEEE ISCA}.

\bibitem[\protect\citeauthoryear{Lin \bgroup et al\mbox.\egroup
  }{2008}]{lin2008automatic}
Lin, Z.; Jiang, X.; Xu, D.; and Zhang, X.
\newblock 2008.
\newblock Automatic protocol format reverse engineering through context-aware
  monitored execution.
\newblock In {\em NDSS}.

\bibitem[\protect\citeauthoryear{Palatucci \bgroup et al\mbox.\egroup
  }{2009}]{palatucci2009zero}
Palatucci, M.; Pomerleau, D.; Hinton, G.~E.; and Mitchell, T.~M.
\newblock 2009.
\newblock Zero-shot learning with semantic output codes.
\newblock In {\em NIPS}.

\bibitem[\protect\citeauthoryear{Pandita \bgroup et al\mbox.\egroup
  }{2013}]{whyper}
Pandita, R.; Xiao, X.; Yang, W.; Enck, W.; and Xie, T.
\newblock 2013.
\newblock Whyper: Towards automating risk assessment of mobile applications.
\newblock In {\em USENIX Security}.

\bibitem[\protect\citeauthoryear{Wang \bgroup et al\mbox.\egroup
  }{2011}]{wang2011inferring}
Wang, Y.; Zhang, Z.; Yao, D.~D.; Qu, B.; and Guo, L.
\newblock 2011.
\newblock Inferring protocol state machine from network traces: a probabilistic
  approach.
\newblock In {\em ACNS}.

\bibitem[\protect\citeauthoryear{Wang \bgroup et al\mbox.\egroup
  }{2013}]{Wang2013}
Wang, J.; Guo, T.; Zhang, P.; and Xiao, Q.
\newblock 2013.
\newblock A model-based behavioral fuzzing approach for network service.
\newblock In {\em IMCCC}.

\bibitem[\protect\citeauthoryear{Witte \bgroup et al\mbox.\egroup }{2008}]{TM}
Witte, R.; Li, Q.; Zhang, Y.; and Rilling, J.
\newblock 2008.
\newblock {Text Mining and Software Engineering: an Integrated Source Code and
  Document Analysis Approach}.
\newblock {\em IET Software}.

\bibitem[\protect\citeauthoryear{Wong \bgroup et al\mbox.\egroup }{2015}]{dase}
Wong, E.; Zhang, L.; Wang, S.; Liu, T.; and Tan, L.
\newblock 2015.
\newblock {DASE:} document-assisted symbolic execution for improving automated
  software testing.
\newblock In {\em ACM/IEEE ICSE}.

\end{thebibliography}
\end{document}